\documentclass[aps,prb,twocolumn,groupedaddress,amsmath,showpacs,floatfix]{revtex4}
\usepackage{graphicx}
\begin{document}
\title{Stability of  $\pi$ junction
configurations  in ferromagnet-superconductor heterostructures}
\author{Klaus Halterman}
\email{klaus.halterman@navy.mil}
\affiliation{Sensor and Signal Sciences Division, Research Department, Naval Air Warfare Center,
China Lake, California 93555}
\author{Oriol T. Valls}
\email {otvalls@umn.edu}
\affiliation{School of Physics and Astronomy and Minnesota Supercomputer
Institute, Minneapolis, Minnesota 55455}
\date{\today}

\begin{abstract}

We investigate the stability of possible order parameter configurations  in 
clean layered heterostructures of the $SFS...FS$ type,
where $S$ is a superconductor and $F$ a ferromagnet. We find that for most
reasonable values of the geometric parameters (layer thicknesses and
number) and of the material parameters (such as magnetic polarization,
wavevector mismatch, and oxide barrier
strength) several solutions
of the 
{\it  self consistent} microscopic equations can coexist, which differ
in the arrangement  
of  the sequence of ``0''
and ``$\pi$'' junction types (that is, with either same or 
opposite 
sign of the
pair potential in adjacent $S$ layers).
The number of such coexisting self consistent solutions increases
with the number of layers.
Studying the 
relative stability of these configurations 
requires an accurate
computation of the small difference in the condensation free energies
of these inhomogeneous systems.
We perform these calculations, starting with  numerical
self consistent solutions of the Bogoliubov-de Gennes equations.
We  present
extensive results for the condensation free energies of the different
possible configurations, obtained by using efficient and accurate numerical 
methods, and discuss their
relative stabilities. 
Results for the experimentally measurable density of states
are also given for
different configurations and
clear differences in the spectra are revealed.
Comprehensive and systematic results as a function
of the relevant parameters for systems
consisting of three and seven layers (one or three junctions) are given, and the
generalization to larger number of layers is discussed. 
\end{abstract}
\maketitle

\section{Introduction\label{introduction}}

A remarkable manifestation of the
macroscopic quantum nature of
superconductivity is seen in the description of the superconducting state
by a complex order parameter
with an associated phase, $\phi$, which is a macroscopic
quantum variable.
For 
composite materials comprised of
multiple superconductor ($S$) layers separated
by nonsuperconducting materials,
the phase difference $\Delta \phi$ between
adjacent $S$ layers becomes a
very relevant quantity.
For the case where  
a nonmagnetic normal metal is
sandwiched between two  superconductors,
it is straightforward to see that the minimum  free
energy configuration  corresponds to
that having a zero phase difference  between the $S$
regions, in the absence of current.
The situation becomes substantially modified
for superconductor-ferromagnet-superconductor ($SFS$) junctions,
where
the presence of the magnetic ($F$)
layers leads to spin-split Andreev\cite{and} states and
to a spatially modulated\cite{ff,lo} order parameter that
can yield a phase difference of
$\Delta \phi=\pi$ between $S$ layers. 
These are the so-called $\pi$ junctions.
Junctions of this type can occur
also in more complicated layered
heterostructures of the $SFSFSF..$ type,
where the relative sign
of the pair potential $\Delta({\bf r})$ can change between adjacent $S$ layers.

Continual improvements
in well controlled deposition
and fabrication techniques have
helped increase the experimental implementations
of systems containing
$\pi$ junctions.
\cite{vavra,bell,jiang,lange,kontos,kontos2,ryazanov,veret,frolov,goldobin} 
Possible applications to devices and to quantum computing,
as well as purely scientific interest, have 
stimulated further interest  
in  these devices.
The pursuit of the $\pi$ state
has consequently generated  
ample supporting data for its existence and properties: 
The observed nonmonotonic behavior in the critical temperature
was found to be consistent with the existence of a $\pi$ state.\cite{jiang}
The domain structure in $F$  is expected to modify the critical
temperature behavior however,
depending on the applied field.\cite{lange}
The ground state of $SFS$ junctions has been recently measured,\cite{kontos} 
and it was found
that  $0$ or $\pi$ coupling existed, depending on the width $d_F$
of the $F$ layer, in agreement
with theoretical expectations. Similarly, damped oscillations in 
the critical current $I_C$
as a function of $d_F$ suggested also a $0$ to $\pi$ transition.\cite{kontos2}
The reported signature in the characteristic $I_C$ curves also indicated a 
crossover from
the $0$ to $\pi$ phase in going from higher to lower 
temperatures.\cite{ryazanov}
Furthermore, the current phase relation was measured,\cite{frolov} 
demonstrating a re-entrant $I_C$ with temperature variation.

A good understanding of the
mechanism and robustness of the $\pi$ state in general
is imperative for the further scientific
and practical development of this area.
The $\pi$ state 
in $SFS$ structures was first investigated  long ago.\cite{bulaevskii}
In general, the exchange field in the ferromagnet shifts the 
different spin bands occupied by the corresponding particle
and hole quasiparticles. This splitting determines
the spatial periodicity of the pair amplitude
in the $F$ layer\cite{dab} and can therefore induce
a crossover from the $0$ state to the $\pi$ state
as the exchange
field $h_0$ varies,\cite{buzdin} or as a function
of $d_F$.
The Josephson critical current was found to be nonzero at the $0$ to $\pi$ 
transition,
as a result from higher harmonics in the current-phase 
relationship.\cite{belzig}
It was found that a coexistence of
stable and metastable states may arise in Josephson junctions,
which
was also attributed to the
existence of higher harmonics.\cite{radovic}
If the magnetization orientation is varied, 
the $\pi$ state may disappear,\cite{bergeret}
in conjunction
with the appearance of a triplet component to the order parameter.
A crossover between $0$ and $\pi$ states by varying the temperature
was explained within the context of a Andreev bound state model
that reproduced experimental findings.\cite{sellier}
In the ballistic limit a transition occurs if
the parameters of the junction are close to the crossover
at zero temperature.\cite{radovic2}
An investigation into the ground states of long (but finite) Josephson junctions
revealed a critical geometrical length scale which separates half-integer
and zero flux states.\cite{zenchuk}
The lengths of the individual junctions were also
found to have important implications, as
a phase modulated state can occur through
a second order phase transition.\cite{buzdin2}

Nevertheless, little work has been done that studies the stability 
of the $\pi$ state for an $SFS$ junction from 
a complete and systematic standpoint, as a function of the
relevant parameters.
This is {\it a fortiori} true for layered systems 
involving more than one junction, each of which
can in principle exist in the 0 or the
$\pi$ state, or
for superlattices. 
There are several reasons for this.
The existence or absence of $\pi$ junction states is intimately
connected to the spatial behavior of $\Delta({\bf r})$
and of the pair amplitude $F({\bf r})$ (which oscillates in
the magnetic layers), and thus
the precise form of these quantities
must be calculated self-consistently, 
so that the resulting $F({\bf r})$ corresponds
to a  minimum in the free energy.
An assumed, non self consistent form for the pair potential, 
typically a piecewise constant
in the $S$ layers,  often deviates 
very substantially from the 
correct self consistent result, and therefore
may often lead to spurious conclusions. Indeed,
the assumed form may in effect
force the form of the final result, thereby
clearly leading to misleading
results.
Self-consistent approaches, despite
their obvious superiority, are too  infrequently
found in the literature
primarily because of the computational expense inherent in the
variational or iterative methods necessary to achieve a solution.
A second problem is that investigation the relative
stability of self-consistent states requires an
accurate calculation of their respective condensation energies,
and that, too, is not an easy problem.

In this paper, we approach in a fully
self consistent manner the question of the stability of
states containing  $\pi$ junctions in $SFSF...S$ multilayer structures.
As has already been shown in trilayers,\cite{hv3} it 
can be the case that, for a given set of geometrical
and material parameters, more than one self
consistent solution  exists, each with a particular 
spatial profile for $\Delta({\bf r})$ involving a junction of either 
the 0 or $\pi$ type. 
It will be demonstrated
below that this is in fact a very common situation in these
heterostructures: one can typically find  several
solutions, all with a negative condensation energy, i.e., they
are all stable with respect to the normal state. Thus a
careful analysis is needed to determine whether each
state is a global or local minimum of the condensation free energy.
This determination is, as we shall see, very difficult
to make from numerical
self-consistent results, because it requires
a very accurate computation of the free energies
of the possible superconducting states,
from which the normal state counterpart must then
be subtracted. This subtraction of
large quantities to obtain a much smaller
one makes the problem  numerically even more challenging than
that of achieving self-consistency, since although the number
of terms involved is the same, the 
numerical accuracy required is much greater.
Until now, this has been seen as a prohibitive numerical obstacle.
Removing this obstacle involves
a careful
analysis and computation of the eigenstates
for each state configuration, that is,
the energy spectrum of the whole system. 

The numerical method we will discuss and implement overcomes these
difficulties, and
therefore enables us to determine the relative
stability of the different states involved, 
as we shall see, for a variety of $F/S$ multilayer structure types, and
broad range of parameters. 
The condensation energies for the several
states found in a fully self consistent manner, are accurately computed as
a function of the relevant parameters.
The material parameters we investigate include,
interfacial scattering, Fermi wave vector mismatch, and
magnetic exchange energy, while the geometrical parameters
are
the superconductor and ferromagnet thicknesses, and total number of layers.
We will see that as the number of  $S$ layers increases, the 
number of possible stable 
junction configurations correspondingly increases.
Our emphasis is on system sizes with  $S$
layers of order of the superconducting coherence length $\xi_0$,
separated by nanoscale magnetic layers.
In order to retain useful information 
that depends
on details at the atomic length scale,
it is necessary to go beyond the various quasiclassical approaches and 
use a microscopic set of equations that does
not average over  spatial variations
of the order of the Fermi wavelength. 
This is  particularly significant in our
multiple layer geometry, where interfering
trajectories
can give important contributions to the quasiparticle spectra,
owing to the specular reflections at the boundaries, and 
normal and Andreev reflections at the interfaces.\cite{ozana}
The influence of these microscopic phenomena is neglected 
in alternative approaches involving averaging over the 
momentum space governing the quasiparticle paths.
Thus, our  starting point is fully microscopic the
Bogoliubov de-Gennes  (BdG) equations,\cite{bdg}
which are a convenient and physically insightful
set of equations that govern
inhomogeneous superconducting systems.
It is also appropriate for the relatively
small heterostructures we are interested in,
to consider the clean limit.

The outline of the paper is as follows. In Sec.~\ref{method}, we 
write down the relevant form of
the real-space BdG equations, and establish 
notation. After introducing an appropriate standing wave
basis, we
develop expressions for the matrix elements needed
in the numerical calculations of the quasiparticle amplitudes and
spectra. 
The iterative algorithm which embodies the self-consistency procedure
is reviewed. 
We then explain how
to use the self consistent pair amplitudes and quasiparticle spectra to 
calculate 
the free energy, as necessary to distinguish among the possible stable 
and metastable states.
In
Sec.~\ref{results}, 
we illustrate for several junction geometries the spatial dependence of the pair
amplitude $F({\bf r})$, which is a direct measure of the proximity 
effect
and gives a physical understanding of the various self
consistent states we find. We display this quantity
as a function of the important physical parameters.
The stability of systems containing
various numbers of $\pi$ junctions is then clarified through a series of
condensation energy calculations that again take into consideration
the material and geometrical parameters mentioned above.
Of importance experimentally  is the local density of states (DOS),
which we also illustrate for certain multilayer configurations.
We find differing signatures for the
possible configurations that should make them discernible in tunneling
spectroscopy experiments.
To conclude, in Sec.~\ref{conclusions} we summarize the results.

\section{\label{method}Method}
\subsection{Basic equations}
In this section we briefly review  the form
that the Bogoliubov-de Gennes (BdG) equations\cite{bdg} take
for the $S/F$ multilayered heterostructures we study.
Additional details can be found in Refs.~\onlinecite{hv1,hv2,hv3}.
The BdG equations are particularly appropriate for the
investigation of the stability of layered configurations
in which the pair amplitude may or may not change  sign
between adjacent superconducting layers. 
These are conventionally called ``$0$'' or ``$\pi$'' junction configurations
respectively.

\begin{figure}
\includegraphics[scale=.5,angle=0]{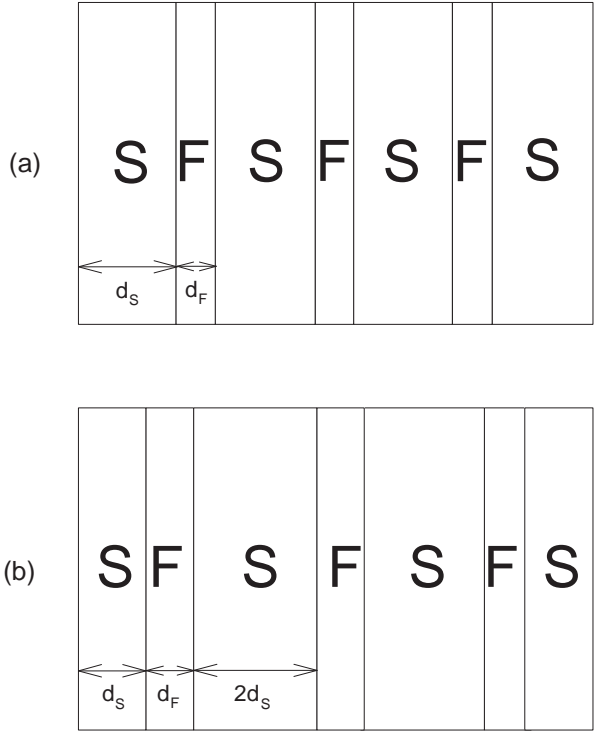}
\caption{\label{schem} Examples of the two types of
multilayer geometries for the  heterostructures examined in this paper.
The system has a total thickness $d$ in the $z$
direction, and the $F$ layers have thickness $d_F$.
The
general patterns shown
hold for structures with an arbitrary odd value of 
the number of layers, $N_L$.
The seven layer case is displayed.
In pattern (a) the thicknesses of each  S layer
is $d_S$, while in (b)  the two outer $S$ layers have thickness $d_S$,
and the  inner ones have  thickness $2 d_S$ (see text). }
\end{figure}

We consider three-dimensional slab-like heterostructures translationally 
invariant
in  the $x-y$ plane, with all spatial variations occurring in the $z$ direction.
The heterostructure consists of superconducting,  $S$, and ferromagnetic,
$F$, layers. Examples are depicted in Fig.~\ref{schem}.
The corresponding coupled equations for the spin-up and spin-down quasiparticle
amplitudes ($u_{n}^{\downarrow},v_{n}^{\uparrow}$) then read
\begin{widetext} 
\begin{subequations}\label{bogo}
\begin{align}
 [-\frac{1}{2m}\frac{\partial^2}{\partial z^2} +\varepsilon_{\perp} 
-E_F(z) + U(z)- h_0(z)]u_n^{\uparrow}(z)+\Delta(z)v_n^{\downarrow}(z)&=\epsilon_n u_n^{\uparrow}(z)\\
-[-\frac{1}{2m}\frac{\partial^2}{\partial z^2} +\varepsilon_{\perp} 
-E_F(z) + U(z)+ h_0(z)]v_n^{\downarrow}(z)+\Delta(z)u_n^{\uparrow}(z)&=\epsilon_n v_n^{\downarrow}(z),
\end{align}
\end{subequations}
\end{widetext}
where $\varepsilon_{\perp}$ is  the kinetic 
energy term corresponding to quasiparticles with momenta transverse to
the $z$ direction, $\epsilon_n$ are the 
energy eigenvalues,  $\Delta(z)$ is the pair potential, and $U(z)$
is the  potential that accounts for scattering at each $F/S$ interface.
An additional set of equations for
$u_{n}^{\downarrow}$ and $v_{n}^{\uparrow}$ can be readily
written down from
symmetry arguments, and thus is suppressed here for brevity.
The form of the ferromagnetic exchange energy $h_0(z)$
is given by the Stoner model, and therefore takes the constant value $h_0$
in the $F$  layers, and zero elsewhere.
Other relevant material parameters are taken into account through
the variable bandwidth
$E_F(z)$. This is
taken to be $E_F(z)=E_{F S}$ in the
$S$ layers, while in the $F$ layers
one has $E_F(z)=E_{FM}$ 
so that in these regions the up and down
bandwidths are respectively $E_{F\uparrow}=E_{FM}+h_0$, and 
$E_{F\downarrow}=E_{FM}-h_0$. The dimensionless parameter
$I$, defined as $I\equiv h_0/E_{FM}$, conveniently characterizes
the magnets' strength, with $I=1$ corresponding to the
half metallic limit. The ratio $\Lambda \equiv E_{FM}/E_{FS}
\equiv (k_{FM}/k_{FS})^2$
describes the mismatch between Fermi wavevectors on the $F$ and $S$
sides, assuming parabolic bands with $k_{FS}$ denoting
the Fermi wave vector in the $S$ regions. 

The spin-splitting effects of the exchange field
coupled with the pairing interaction in the $S$ regions,
results in a nontrivial 
spatial dependence of the pair potential, which is further
compounded by the normal and Andreev scattering events that 
occur at the  multiple $S/F$ interfaces. When
these complexities are  taken into account, one
generally cannot  assume any explicit form for  $\Delta(z)$
{\it a priori}. Thus,
when solving Eqs. (\ref{bogo}), the pair potential must be calculated in a self
consistent manner by an appropriate sum over states:
\begin{widetext}
\begin{equation}  
\label{del2} 
\Delta(z) = \frac{\pi g(z) N(0)}{k_{FS} d}\sum_{\epsilon_n\leq \omega_D}\int{d\varepsilon_\perp}
\left[u_n^\uparrow(z)v^\downarrow_n (z)+
u_n^\downarrow(z)v^\uparrow_n (z)\right]\tanh(\epsilon_n/2T),
\end{equation} 
\end{widetext}
where  $N(0)$ is the DOS per spin of the 
superconductor
in the normal state, $d$ is the total
system size in the $z$ direction, $T$ is the temperature, $\omega_D$ is the 
cutoff ``Debye'' energy of the pairing interaction,
and $g(z)$ is the effective  coupling, which we take to be a constant
$g$ within the  superconductor regions and zero elsewhere.
 
The presence of interfacial scattering is expected to
modify the proximity effect.
We assume that every $S/F$ interface induces the same scattering potential,
which we take of a delta function form: 
\begin{equation}
U(z_l,z)=H \delta(z-z_l)
\label{inter}
\end{equation}
where $z_l$ is the location
of the interface and $H$ is the scattering parameter. It is convenient to
use the dimensionless parameter $H_B\equiv mH/k_{FS}$ to characterize
the interfacial scattering strength.

An appropriate choice of basis allows
Eqs.~(\ref{bogo})
to be 
transformed into a finite $2N \times 2N$ dimensional matrix eigenvalue 
problem in wave vector space:
\begin{equation}
\label{mbogo}
\begin{bmatrix} H^+&D \\
D & H^- \end{bmatrix}
\Psi_n
=
\epsilon_n
\,\Psi_n,
\end{equation}
where 
$\Psi_n^T =
(u^{\uparrow}_{n1},\ldots,u^{\uparrow}_{nN},v^{\downarrow}_{n1},
\ldots,v^{\downarrow}_{nN})$, are the expansion coefficients
associated with
the set of orthonormal basis vectors,
$u^{\uparrow}_n(z)=\sqrt{{2}/{d}}\sum_{q=1}^N u^{\uparrow}_{n q}\sin(k_q z)$, and
$v^{\downarrow}_n(z)=\sqrt{{2}/{d}}\sum_{q=1}^N v^{\downarrow}_{n q}
\sin(k_q z)$. The longitudinal momentum index $k_q$
is quantized via
$k_q = {q/\pi}{d}$, where
$q$ is a positive integer.
The label $n$ encompasses the index $q$
and the value of $\varepsilon_\perp$. 
The finite range of the pairing interaction $\omega_D$, implies
that $N$ is finite.
In our layered geometry submatrices corresponding to different values of
$\varepsilon_\perp$ are decoupled from each other, so one considers 
matrices labeled by the $q$ index, for each relevant value of 
$\varepsilon_\perp$. 
The  matrix elements in Eq. (\ref{mbogo}) depend in general on
the  geometry under consideration, and are
given for two specific cases in the subsections below.

\subsection{Identical superconducting  layers }

The first type of structure we consider is one consisting of alternating
$S$ and $F$ layers, each of width $d_S$
and $d_F$ respectively. 
This  geometry is shown in Fig. \ref{schem}(a) for the 
particular case of $N_L=7$. 
For a given total number of layers (superconducting plus 
magnetic) $N_L$, we have in this case for the interfacial scattering:
\begin{equation}
U(z)=\sum_{i=1}^{{(N_L-1)}/{2}}[U(i(d_S+d_F),z)+U((i(d_S+d_F)-d_F)),z]
\end{equation}
where $U(z_l,z)$ is given in Eq.~(\ref{inter}).
The matrix elements $H^+_{q q'}$ and $H^-_{q q'}$
in Eq.~(\ref{mbogo}) 
are compactly written for this geometry  as
\begin{widetext}
\begin{subequations}
\begin{align}
H^{+}_{q q}&=\frac{k^2_q}{2m} + \varepsilon_{\perp}+\frac{2H}{d}\left[\frac{N_L-1}{2}-A(2q)\right]
-E_{F \uparrow}\left[\frac{d_F}{d}\left(\frac{N_L-1}{2}\right)+B(2q)\right] \nonumber \\ 
&-E_{FS}\left[\frac{d_S}{d}\left(\frac{N_L+1}{2}\right)-B(2q)\right],\\
H^{+}_{q q'}&=\frac{2H}{d}[A(q-q')-A(q+q')]+[E_{F \uparrow}-E_{FS}][B(q-q')-B(q+q')],& q\neq q',
\end{align}
\end{subequations}
where,
\begin{subequations}
\begin{align}
A(q)&=\cos\left(\frac{d_F \pi q}{2 d}\right) \sum_{i=1}^{(N_L-1)/2}
\cos\left[\frac{\pi q}{2d}(2 d_S i + d_F(2i-1))\right], \\
B(q)&=\frac{2 \sin\left(\frac{d_S \pi q}{2 d}\right)}{\pi q} \sum_{i=1}^{(N_L+1)/2}
\cos\left[\frac{\pi q}{2d}(d_S (2i-1) + 2d_F(i-1))\right].
\end{align}
\end{subequations}
The matrix elements  $H^{-}_{q q'}$ are similarly expressed in term of
the coefficients $A(q)$ and $B(q)$,
\begin{subequations}
\begin{align}
H^{-}_{q q}&=-\frac{k^2_q}{2m} - \varepsilon_{\perp}-\frac{2H}{d}\left[\frac{N_L-1}{2}-A(2q)\right]
+E_{F \downarrow}\left[\frac{d_F}{d}\left(\frac{N_L-1}{2}\right)+B(2q)\right] \nonumber \\ 
&+E_{FS}\left[\frac{d_S}{d}\left(\frac{N_L+1}{2}\right)-B(2q)\right],\\
H^{-}_{q q'}&=-\frac{2H}{d}[A(q-q')-A(q+q')]-[E_{F \downarrow}-E_{FS}][B(q-q')-B(q+q')],& q\neq q'.
\end{align}
\end{subequations}
\end{widetext}

The  $D_{q q'}$  in the off-diagonal part
of the left side of Eq.~(\ref{mbogo})
arise from an integral over $\Delta(z)$, which
scatters a quasiparticle of a given spin into  a quasihole of opposite spin.
One has:
\begin{equation}
D_{q q'}=\frac{2}{d}\int_0^d dz \sin(k_q z)\Delta(z) \sin (k_q' z).
\end{equation}
It is straightforward to write also\cite{hv2,hv3} the self consistency 
equation in terms of matrix elements. 

\subsection{Half-width superconducting outer layers }
The previous subsection outlined the details needed to 
arrive at the matrix elements 
when the $S$ layers
are of the same width $d_S$. We are also interested in investigating structures
where the inner $S$ layers are twice as thick ($2d_S$) as the outer ones 
(see Fig. \ref{schem}(b)) while the $F$
layers remain all of the same width. This case is of interest
because, roughly speaking, the inner layers, being
between ferromagnets, should experience about twice the
pair-breaking effects of the exchange field than do the outer ones.
Therefore, the results might depend on $N_L$ more systematically,
particularly for relatively small $N_L$, if the witdth of the outer
layers is halved. This has been found to be the case in
some studies\cite{prosh} of the transition temperature in thin
layered systems.   

A slight modification to the previous results yields the following form of the
scattering potential $U(z)$,
\begin{widetext}
\begin{equation}
U(z)=\sum_{i=1}^{{(N_L-1)}/{2}}[U((i(2d_S+d_F)-d_S),z)+U((i(2 d_S+d_F)-d_S-d_F),z)]
\end{equation}
The matrix elements  $H^{+}_{q q'}$ 
are 
now expressed as
\begin{subequations}
\begin{align}
 H^{+}_{q q'}&=
\frac{2H}{d}[A(q-q')-A(q+q')]+[E_{F \uparrow}-E_{FS}][B(q+q')-B(q-q')],\qquad q\neq q'\\
H^{+}_{q q}&=\frac{k^2_q}{2m} + \varepsilon_{\perp}+\frac{2H}{d}\left[\frac{N_L-1}{2}-A(2q)\right]
-E_{F \uparrow}\left[\frac{d_F}{d}\left(\frac{N_L-1}{2}\right)-B(2q)\right] \nonumber \\
&-E_{FS}\left[\frac{d_S}{d}\left(N_L-1\right)+B(2q)\right],
\end{align}
\end{subequations}
where  the coefficients $A(q)$ and $B(q)$ now read,
\begin{align}
A(q)=\cos\left(\frac{d_F \pi q}{2 d}\right) \sum_{i=1}^{(N_L-1)/2}
\cos\left[\frac{\pi q}{2d}(2i-1)(2d_S+d_F)\right], \qquad
B(q)=\frac{2A(q)}{\pi q} \tan\left(\frac{d_F \pi q}{2 d}\right).
\end{align}
In a similar manner, The matrix elements  $H^{-}_{q q'}$ are written as
\begin{subequations}
\begin{align}
H^{-}_{q q}&=-\frac{k^2_q}{2m} - \varepsilon_{\perp}-\frac{2H}{d}\left[\frac{N_L-1}{2}-A(2q)\right]
+E_{F \downarrow}\left[\frac{d_F}{d}\left(\frac{N_L-1}{2}\right)-B(2q)\right] \nonumber \\
&+E_{FS}\left[\frac{d_S}{d}\left(N_L-1\right)+B(2q)\right], \\
 H^{-}_{q q'}&=
-\frac{2H}{d}[A(q-q')-A(q+q')]-[E_{F \downarrow}-E_{FS}][B(q+q')-B(q-q')],\qquad q\neq q' \nonumber \\
\end{align}
\end{subequations}
\end{widetext}

The matrix elements of $D$ are as in the previous subsection, and the
self consistent equation can be similarly rewritten.

\subsection{Spectroscopy}

Experimentally
accessible information regarding the quasiparticle spectra is contained in
the local density of
one particle excitations in the system. 
The local density of states (LDOS) for
each  spin orientation is given by 
\begin{equation}\label{dos}
{N}_\sigma(z,\epsilon) 
=-\sum_{n}
\Bigl\lbrace[u^\sigma_n(z)]^2
 f'(\epsilon-\epsilon_n) 
+[v^\sigma_n(z)]^2
 f'(\epsilon+\epsilon_n)\Bigr\rbrace.  
\end{equation}
where $\sigma=\uparrow,\downarrow$ and
$f'(\epsilon) = \partial f/\partial \epsilon$ is the derivative
of the Fermi function.

As discussed in the Introduction, the 
condensation free energies of the different 
self consistent solutions found
must be compared\cite{hv3} in order to find the most
stable configuration, as opposed to those that are metastable.
While for homogeneous systems this  quantity is found
in standard textbooks,\cite{tinkham,fw}.
the case of an inhomogeneous system is more complicated.
We will use the convenient expression found in Ref.~\onlinecite{kos}
for the free energy ${\cal F}$:
\begin{equation}
{\cal F} = -2 T \sum_n {\ln\left[2 \cosh\left(\frac{\epsilon_n}{2T}\right)\right]}+
\int^d_0 dz \frac{|\Delta(z)|^2}{g},
\label{free}
\end{equation}
where the sum can be taken
over states of energy less than $\omega_D$.
For a uniform
system the above
expression properly reduces to the standard
textbook result.\cite{fw} The corresponding condensation free energy
(or, at $T=0$, the condensation
energy) is obtained by subtracting the corresponding normal state quantity,
as discussed below.  
Thus, in principle, only the results for
$\Delta(z)$ and the excitation spectra are needed to calculate the free energy.
As  pointed out in 
Ref.~\onlinecite{hv3}, a  numerical computation of the
condensation energies that is accurate enough to allow comparison between states
of different types
requires great care and accuracy. Details 
will be given in the next section. 

\section{\label{results}Results}

As explained in the Introduction, the chief objective of this work
is to study the relative stability of the different states that
are obtained through self-consistent solution of the BdG equations
for this geometry. These solutions differ in the nature of the 
junctions. Each  junction between two consecutive S layers can
be of the ``0'' type (with the order parameter in both S layers
having the same phase) or of the ``$\pi$'' type (opposite phase). As
the number of layers, and junctions, increases, the number of 
order parameter (or junction) configurations 
which are in principle possible increases also. As we shall see,
for any set of parameter values (geometrical and material) not all
of the possible configurations are realized: some do not correspond
to free energy local minima. Among those that do, 
the one (except for accidental degeneracies) which is
the absolute stable minimum must be determined, the other
ones being metastable.
We will discuss these stability questions as a function of the
material and geometrical properties, as represented
by dimensionless parameters as we shall now
discuss. 

\subsection {General considerations}

Three material parameters are found
to be very important: one is obviously the magnet strength
$I$. We will vary this parameter in the range from zero to one,
that is, from nonmagnetic to half-metallic. The second
is the wavevector mismatch characterized by
$\Lambda \equiv (k_{FM}/k_{FS})^2$.
The importance of this parameter can be understood by considering
that, even in the non self consistent limit, the different
amplitudes for ordinary and Andreev scattering depend strongly on
the wavevectors involved, as it follows from elementary considerations.
We will vary $\Lambda$ in the range from unity (no mismatch)
down to $1/10$. We have not considered values larger than unity
as these are in practice infrequent. The third important dimensionless
parameter is the barrier height $H_B$   defined below Eq.~(\ref{inter}).
This we will vary from zero to unity, at which
value the $S$ layers become, as we shall
see, close to being decoupled.  We will keep the superconducting
correlation length fixed at $k_{FS} \xi_0 \equiv \Xi_0 =100$. This quantity 
sets the length scale for the superconductivity and therefore can
be kept fixed, recalling only that, to study the $d_S$ dependence, one needs
to consider the value of $d_S/\xi_0$. Finally the cutoff frequency $\omega_D$
can be kept fixed (we set $\omega_D=0.04 E_{FS}$) since it sets the 
overall energy scale
and we are interested in relative shifts.
The dimensionless coupling constant $g N(0)$ can be derived
from these quantities using standard relations.
In this study we will focus on very low temperatures
limit, fixing $T$ to 
$T=0.01 T_C^0$ (where $T_C^0$ is the bulk transition value).
The geometrical parameters are obviously the number of layers $N_L$, 
and the thicknesses $d_F$ 
and $d_S$. We will consider two examples of the first: $N_L=3$ and $N_L=7$.
For the larger value we will study both of the geometries in panels (a)
and (b) of Fig.~\ref{schem}. The thicknesses, usually expressed in terms
of the dimensionless quantities $D_S \equiv k_{FS} d_S$ and 
$D_F \equiv k_{FS} d_F$,  will  be varied over rather extended ranges.

As we study the effect of each one of these parameters by varying it in
the appropriate range, we will be holding the others constant
at a certain value. Unless otherwise indicated, the values of
the parameters held constant will take the
``default'' values $I=0.2$, $D_S=100=\Xi_0$, $D_F=10$, $\Lambda=1$,
and $H_B=0$. One important derived length is  
$(k_\uparrow-k_\downarrow)^{-1}$, where $k_\uparrow$ and $k_ \downarrow$
are the Fermi wavevectors of the up and down magnetic bands. As is
well known\cite{dab, hv1}, this quantity determines the approximate  spatial
oscillations of the pair amplitude in the magnet. In terms of
the quantities $I$ and $\Lambda$ we can define:
\begin{equation}
\label{xi2}
\Xi_2 \equiv k_{FS}(k_\uparrow-k_\downarrow)^{-1}= \frac{1}{\Lambda^{1/2}}
\frac{1}{\sqrt{(1+I)}-\sqrt{(1-I)}}
\end{equation}
At $I=0.2$ and $\Lambda=1$ one has $\Xi_2=4.97$ increasing to $15.7$ at
$\Lambda=0.1$. This motivates our default choice $D_F=10$.

The numerical algorithm used in our self consistent calculations
follows closely that of  previous developed codes used in simpler
geometries.\cite{hv2,hv3} There are 
however some extra complexities that arise for
the larger multilayered structures studied here, and from
the increased number of self consistent states to be analyzed.
As usual, one must
assume an initial particular form for the pair potential, to
start the iteration process. This
permits a straightforward diagonalization of the matrix
given in Eq.~(\ref{mbogo}) for a given set of geometrical and material
parameters,
for each value of the transverse energy $\varepsilon_\perp$.
The initial guess of $\Delta(z)$ is always chosen as a piecewise constant
$\pm \Delta_0$, where $\Delta_0$ is the zero temperature
bulk gap, and the signs depend on the possible configuration
being investigated (see below). 
Self consistency is deemed to have been achieved
when the difference between two successive  $\Delta(z)$'s
is less than $10^{-5}\Delta_0$ at every value of $z$.
The minimum number of $\varepsilon_\perp$
variables needed for self consistency is around $N_\perp=500$ 
different values of $\varepsilon_\perp$.
In practice however, use of a value
close to this minimum is insufficient to produce
smooth results for 
the local DOS.
Therefore, we first calculate  $\Delta(z)$
self consistently using  $N_\perp=500$, after which iteration is continued
with $N_\perp$ increased by a factor of ten.
The computed spectra then summed according to  Eq.~(\ref{dos})
are smooth and further increases
in $N_\perp$ produce no discernible change in the outcome.
This two-step procedure leads to considerable
savings in computer time.  
The convergence properties and net computational expense obviously depend
also on the 
dimension of the matrix to be diagonalized,  dictated by the number
of basis functions, $N$, which scales linearly
with system size $k_{FS}d$. Thus, the 
computational
time then increases approximately as $(k_{FS}d)^2$, which 
can be a formidable issue.
For the largest structures considered in this paper, the resultant matrix 
dimension
is then $2N\times2N \simeq 2000 \times 2000$ for each $\varepsilon_\perp$.

The number of iterations needed for self consistency depends on the
initial guess. If the self consistent state turns out to be composed of
junctions of the same types as
the initial guess, as specified by the signs in the $\pm \Delta_0$,
then forty or fifty iterations  are usually sufficient. 
But
a crucial point is that, as we shall see, not all of the initial
junction configurations 
lead to self consistent solutions of the same type. 
Since the self-consistency condition is derived from a free energy minimization
condition, this means that  
some  of the initial guesses do not
lead to minima in the free energy with the same junction
configuration as the initial guess, and thus that locally stable states
of that type do not exist, for the particular parameter values studied.
In this case, the number of iterations required to get a
self consistent solution increases dramatically, since the pair 
potential has to reconfigure itself into a much different profile.
The number of iterations in those cases can exceed 400.

The computation of the
condensation free energies of the self
consistent states found is still a difficult task, even
after the spectra are computed: 
considering for illustration the  $T=0$ limit, recall\cite{tinkham} 
that the bulk
condensation energy, given by
\begin{equation}
E_0^b=-(1/2) N(0) \Delta_0^2
\label{bulke}
\end{equation}
represents a fraction of the energy
of the relevant electrons of order $(\Delta/\omega_D)^2$, a quantity of order
$10^{-2}$ or less. The condensation free energies of the inhomogeneous
states under discussion are usually much smaller (by about an order of 
magnitude as we shall see)
than that of the bulk. Distinguishing among competing states by comparing
their often similar condensation free energies, requires computing these 
individual
condensation energies to very high precision. 
This  task is numerically difficult. Here the advantages of  
the expression from Ref.~\onlinecite{kos}
which we use (see Eq.~(\ref{free})) are obvious.
Only the energies are explicitly
needed in the sums and integrals 
on the right side, the influence of the eigenfunctions being only indirectly
felt through the relatively smooth function $\Delta(z)$. The required
quantities are obtained with sufficient precision, as we shall demonstrate,
from our numerical calculations.
Some of the
different but equivalent expressions found
in the literature for the condensation energies or
free energies\cite{fw,ks}, contain explicit sums over eigenfunctions,
which can lead to cumulative errors. Also, the approximate formula used
for the condensation energy of trilayers in Ref.~\onlinecite{hv3}, consisting  in
replacing, in Eq.~(\ref{bulke}),  the pair amplitude with its spatial
average, 
while plausible,
is difficult to validate in general, and
becomes inaccurate for systems involving a larger number
of thinner layers, 
particularly with $S$ widths of order of $\xi_0$.

Using this procedure, the task then becomes feasible but not trivial as it
involves a sum over more than $10^6$ terms. Also, to obtain the
condensation energy one still has to subtract from  the superconducting
${\cal F}$ the corresponding normal quantity ${\cal F_N}$. It is from 
subtracting these larger quantities that the much smaller condensation
free energy is obtained. 
The normal state energy spectra used to evaluate ${\cal F_N}$ are  computed
by taking $D=0$ in Eq.~(\ref{mbogo}), and diagonalizing the resulting matrix. 
The presence of interfacial scattering and mismatch in the Fermi wavevectors
still introduces off-diagonal matrix elements.
In performing the subtraction, care must be taken (as in fact
is also the case in the bulk analytic case) to include exactly the
same number of states in both calculations, rather than loosely using the
same energy cutoff: no states overall
are created or lost by the phase transition. With
all of these precautions, results of sufficient precision and smoothness are
obtained. Results obtained with alternative equivalent\cite{bkk}
formulas are consistent but noisier. We have  verified that in the 
limiting case of large
superconducting slabs our procedure reproduces the analytic bulk results.

\subsection{Pair amplitude}

\begin{figure}
\includegraphics[scale=1.0,angle=0]{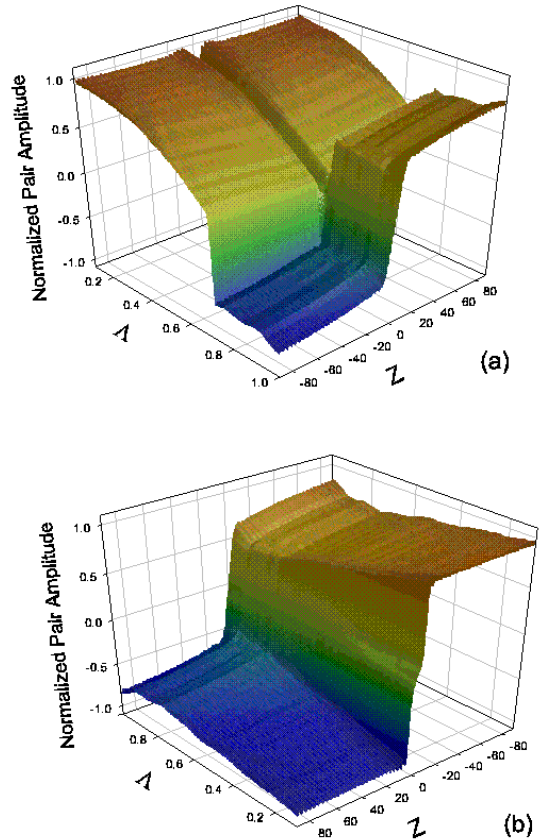}
\caption{\label{fig2} (Color online). The pair amplitude $F(Z)$, normalized
to $\Delta_0$, for a three 
layer $SFS$
structure, plotted as a function of $Z\equiv k_{FS} z$ and of the mismatch
parameter $\Lambda$, at $H_B=0$. 
The direction of the $\Lambda$ scale is different in 
the top and bottom panels.
The $Z=0$ point is at the center of the structure.
We have $D_S \equiv k_{FS} d_S=100$ and $D_F \equiv k_{FS} d_F=10$. Panel (a)
corresponds to self consistent results obtained
with an initial guess where the
junction is of the ``0'' type and panel (b) with
a ``$\pi$'' type. 
In the latter case, the solution found is always of the
$\pi$ type, but in the former a solution of the 0 type
is obtained only for large mismatch (small $\Lambda$).
We have $I=0.2$ and $T=0.01 T_c^0$ here. See text for discussion. 
 }
\end{figure}

We begin by presenting some results for the pair amplitude $F(z)$, showing
how this quantity varies as a function of the interface scattering
parameter $H_B$ and the Fermi wavevector mismatch $\Lambda$.  This is best
done by means of three dimensional plots. In the first of these,
Fig.~\ref{fig2}, we show the pair amplitude (normalized to $\Delta_0$)
for a three layer $SFS$
system, with $D_F=10$ and $D_S=100$, 
as a function of position and of mismatch parameter $\Lambda$, at $H_B=0$.
The top panel shows the results of attempting to find a solution of the
``0'' type by starting the iteration process
with an initial guess of that form. Clearly such an attempt fails at small 
mismatch
($\Lambda \gtrsim 0.7$) indicating that a solution of this type is 
then unstable.
At  larger mismatch, a $0$ state solution 
is found. Panel (b) shows solutions
obtained by starting with an initial guess of the ``$\pi$'' type: a 
self consistent solution
of this type is then always obtained (note that the
$\Lambda$ scales run in different directions
in panels (a) and (b)). For $\Lambda \gtrsim 0.7$, the solutions
of course turn out to be the same as found in panel (a) in the same range. Thus, we
see that for small mismatch there is only one self consistent solution, 
which is of the $\pi$ type, while when the
mismatch is large
there are two competing solutions and their relative
stability becomes an issue.  

\begin{figure}
\includegraphics[scale=.8,angle=0]{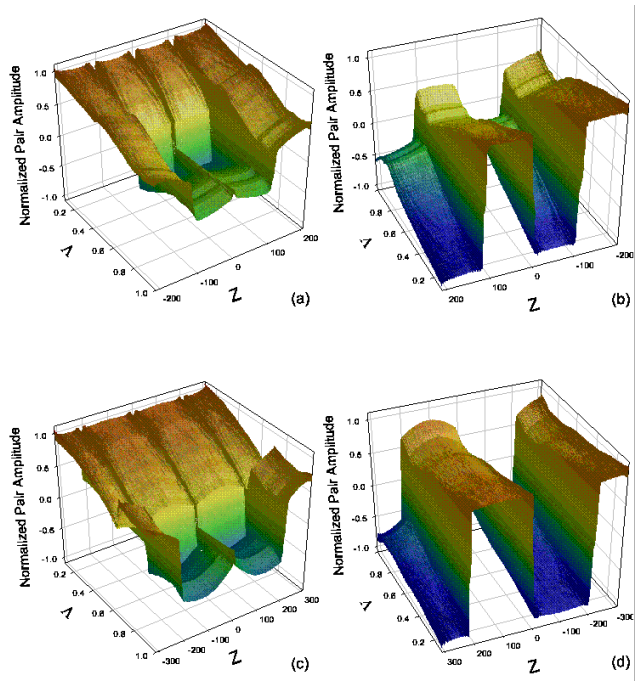}
\caption{\label{fig3} (Color online). The normalized
pair amplitude $F(Z)$ for a 
seven layer $SFSFSFS$
structure, plotted as in Fig.~\ref{fig2} for the same parameter
values. In panels (a) and (b), the thickness of all S layers is the same,
while in panels (c) and (d) the thickness of the two inner S layers is
doubled to $2 D_S =200$ (see Fig.~\ref{schem}). Panels (a) and (c)
correspond to an initial guess of the ``000'' type and panels (b) and (d) to
a ``$\pi\pi\pi$'' type, (see text). The configuration
of the plotted self consistent results  can be ``000'', ``$\pi\pi\pi$''
or ``$\pi0\pi$'' as explained in the text.
 }
\end{figure}

\begin{figure}
\includegraphics[scale=1.,angle=0]{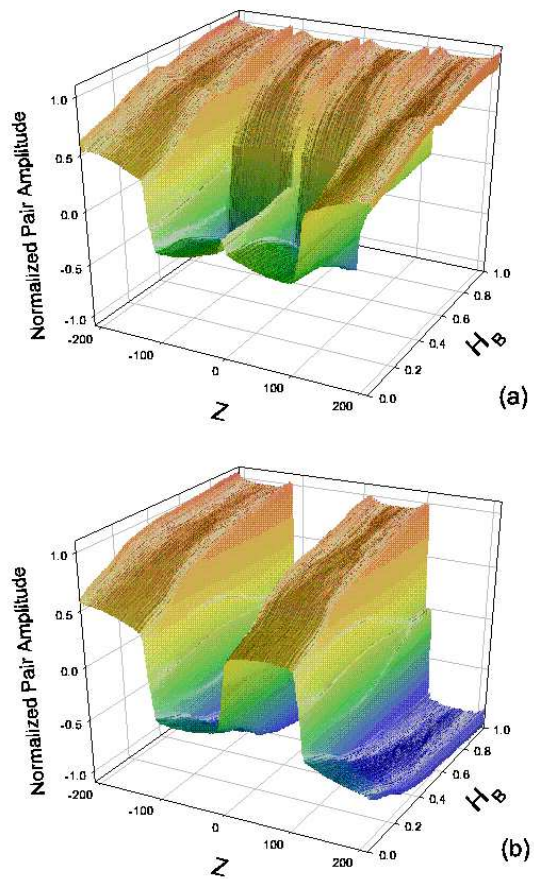}
\caption{\label{fig4}(Color online). The pair amplitude $F(Z)$ for a seven layer
structure, plotted for the same parameter values and conventions
as in the upper panels of Fig.~\ref{fig2}, but as a function of the dimensionless
barrier height $H_B$ (at $\Lambda=1$), rather than of the mismatch.  
}
\end{figure}

We now  turn to seven layer $SFSFSFS$ structures.
In classifying the different possible configurations,
it is convenient to establish a notation that envisions the
seven layer geometry as consisting of
three adjacent $SFS$ junctions. Thus, up to
a trivial reversal, we can then
denote as ``$000$" the structure when adjacent $S$ layers always have the same 
sign of $\Delta(z)$,  and as ``$\pi\pi\pi$" the structure
where successive
$S$ layers alternate in sign.
There are also two other distinct symmetric states: one in which $\Delta(z)$
has the same sign in the first
two $S$ layers, and in the last two it has the opposite sign,
(this is labeled as the ``$0\pi0$" configuration), and the other 
corresponding to
the two outer $S$ layers having the same sign for $\Delta(z)$, 
opposite to that of the two inner $S$ layers: these are 
referred to as ``$\pi0\pi$" structures in this notation.
We will focus our study on these symmetric configurations.
Asymmetric configurations corresponding 
in our notation to the $\pi00$, and $\pi\pi0$ states
are not forbidden, but occur very rarely and will
be
addressed only as need may arise.
In Fig.~\ref{fig3} we repeat the plots in Fig.~\ref{fig2} for seven
layer structures.
We include the cases in which all $S$ layers are of the same thickness 
(top two panels)
and the case where the thickness of the two inner $S$ layers is doubled 
(bottom panels, see
Fig.~\ref{schem}). 
In panels (a) and (c), the initial guess is of the
$000$ type, while in (b) and (d) it is of the $\pi\pi\pi$ type.
In the  case of identical $S$ layer widths, 
we see  that  a $000$ guess 
(Fig.~\ref{fig3}(a))
yields a self consistent state  of the same $000$ form only for larger
mismatch, $\Lambda \lesssim 0.5$, while for smaller  mismatch the
configuration obtained
is
clearly of the $\pi0\pi$ form. Thus there is a value of $\Lambda$
where two self consistent solutions  cross over. However,
a $\pi\pi\pi$ guess (Fig.~\ref{fig3}(b))
results in a self consistent $\pi\pi\pi$ configuration for the whole 
$\Lambda$ range.
Thus, there is a clear competition between {\it
at least} these three observed states,
resulting from multiple minima of the free energy.
Solutions of the $0\pi0$ type are not displayed in this Figure
but they will be discussed below. 
In the two bottom panels we see the same effects when
the thickness of the inner S layers is doubled. As
explained above, this describes a  more balanced situation, since the inner
layers have magnetic neighbors on both sides. 
It is evident from the figure that the pairing correlations are
increased in the $S$ layers. In Fig. \ref{fig3}(c), there is also a 
noticeable
shift in the crossover
point separating the $000$ and $\pi0\pi$ self consistent states,
occurring now at smaller mismatch $\Lambda \approx 0.7$.
Again we find the competition between 
the various states  extends through the entire range
of $\Lambda$ considered. 

\begin{figure}[t!]
\includegraphics[scale=1.0,angle=0]{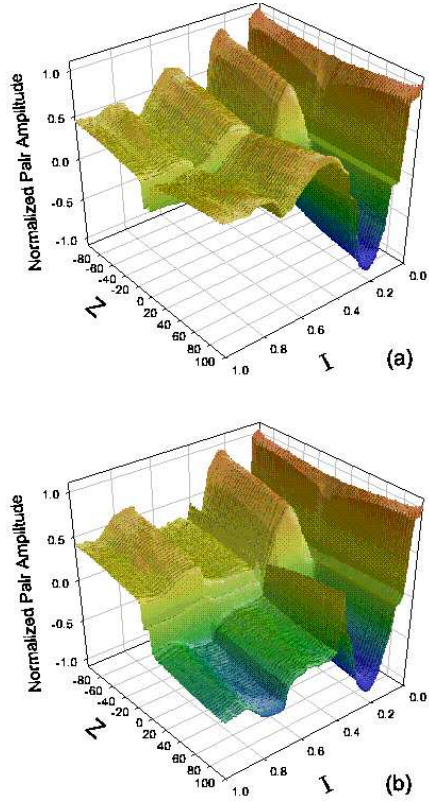}
\caption{\label{fig5} (Color online). The pair amplitude $F(Z)$ for a three 
layer
structure, plotted for the same parameter values and conventions
as in Fig.~\ref{fig2}, but as a function of the dimensionless
magnetic polarization $I$, at $\Lambda=1$ and $H_B=0$. 
 }
\end{figure}

Next we consider in Fig.~\ref{fig4} the influence of the barrier height
as represented by the dimensionless parameter $H_B$. This figure is
completely analogous to Fig.~\ref{fig3}(a) and (b), except for the substitution of
$H_B$ for $\Lambda$, which is set
to unity (no mismatch). We find, by examining and comparing the panels that
two solutions of the $\pi\pi\pi$ and $\pi0\pi$ type exist for small barrier 
heights ($H_B \lesssim 0.5$),
but when $H_B$ becomes larger the $\pi\pi\pi$ and $000$ states then compete. 
Another trend which
one can clearly discern is that the absolute values of $F(Z)$ in the middle
of the S layers increase with $H_B$. This makes  sense, as at larger
barrier heights the layers become more isolated from each other, and the
proximity effects must  in general weaken. 

To conclude this discussion of $F(z)$ we display in 
Fig.~\ref{fig5}, the pair amplitude for the same three layer system as in
Fig.~\ref{fig2}, as a function of position and
of the magnetic exchange parameter $I$ in its entire range, without
a barrier or mismatch. Careful examination of the two panels
reveals a rather intricate situation:  a solution of the $\pi$ type
exists nearly in the entire  $I$ range (see Fig.~\ref{fig5}(b)), the 
exception being at very small
$I$, where the magnetism becomes
weaker and, as one would expect, only the $0$ state solution is found.  This
requires  small values of $I$,  $I \lesssim 0.1$ however.
One can see that in a small neighborhood of $I \approx 0.1$, as the
$0$ state transitions to the $\pi$ state, the pair amplitude is 
small throughout the structure. This may correspond to a marked dip
in the transition temperature.
At larger
values of $I$, the two types of solutions coexist, but there is a range
around $I=0.2$ where clearly there is only one.

These results, which include for brevity only a very small
subset of those obtained, are 
sufficient, we believe, to persuade the reader that
although for some parameter values a unique self consistent solution
exists, this is comparatively rare, and that in general
several  solutions
of differing symmetry types can be found. These self consistent solutions
correspond to local free energy minima: they
are at least metastable. Furthermore, it is clear that
the uniqueness or multiplicity of solutions depends in a complicated way
not only on the geometry, but also on the specific material parameter
values.  

\begin{figure}[t]
\includegraphics[scale=.5,angle=0]{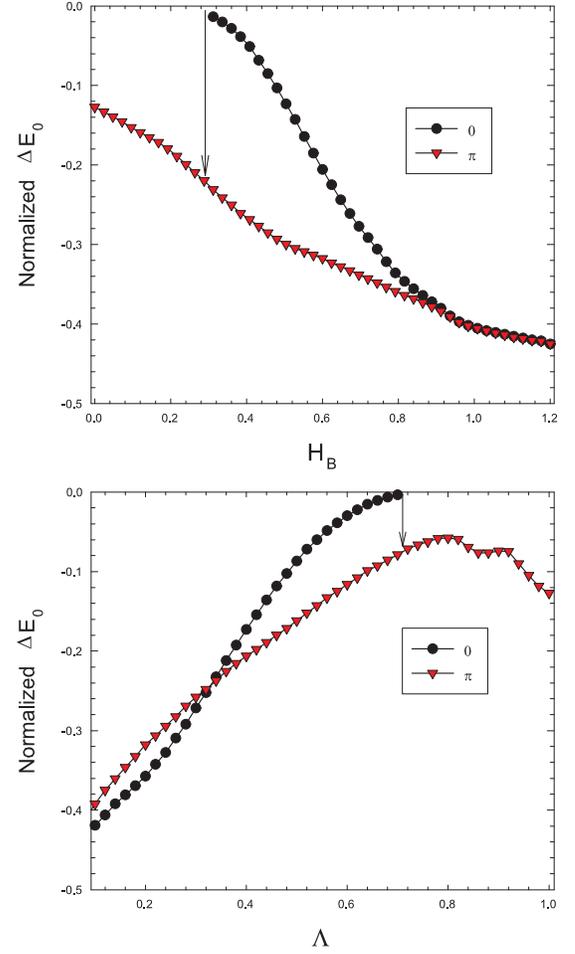}
\caption{\label{fig6} (Color online). The  
normalized condensation free energy $\Delta E_0$ (see text) of a three layer
$SFS$ structure, plotted as a function of barrier height (top)
and mismatch parameter $\Lambda$ (bottom) for self consistent
states of both the 0
and $\pi$  types, as indicated. All other material parameters, geometrical
values, and temperature, are as in Fig.~\ref{fig2}.
The vertical arrow marks the end, as either $H_ B$ increases (top) or
$\Lambda$ increases (bottom), of the region of stability of the 0 state
in this case.
 }
\end{figure}

\subsection {Condensation Free Energy: Stability}

One must, in view of the results
in the previous subsection, find a way to determine in each case the 
relative stability of each configuration and the global free energy
minimum. This is achieved by computing the free energy of the
several self consistent states, using the accurate numerical
procedures explained earlier in this
Section. Results for this quantity, which at the low temperature studied is
essentially the same as the condensation energy, are given in the
next figures below.
The quantity plotted in these figures, which  we
call the normalized $\Delta E_0$,
is the condensation free energy (as calculated from 
Eq.~(\ref{free}) after normal state subtraction)  
normalized to $N(0) \Delta_0^2$, which is twice the zero temperature
bulk value  (see Eq.~\ref{bulke}). Therefore,
at the low temperatures studied, the bulk  uniform
value of the quantity plotted 
is very close to $-(1/2)$.

\begin{figure}[t]
\includegraphics[scale=.5,angle=0]{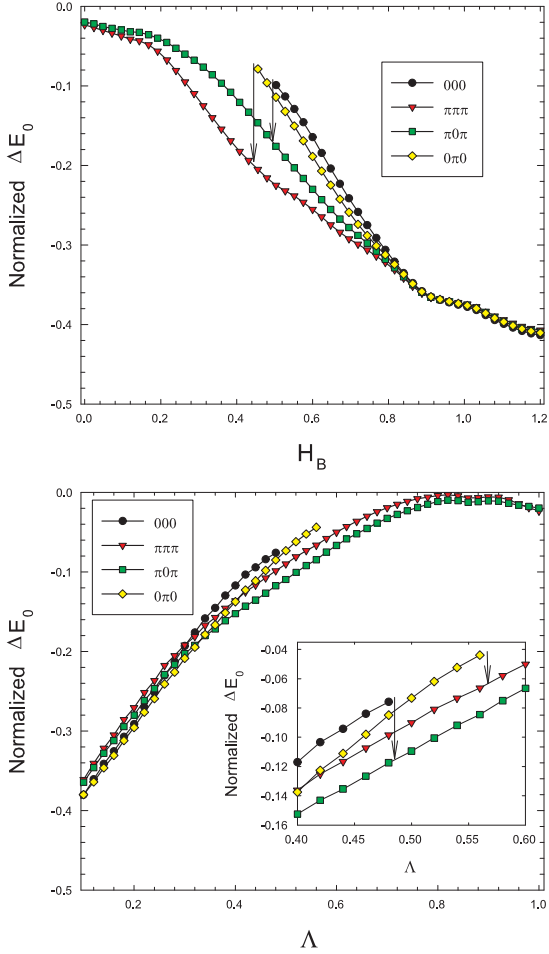}
\caption{\label{fig7} (Color online). The  normalized $\Delta E_0$ 
for a seven layer
$SFSFSFS$ structure plotted as a function of barrier height (top)
and mismatch parameter $\Lambda$ (bottom) for self consistent states 
of the  types indicated (see text for
explanation). Material parameter and geometrical
values are as in previous Figures, and all $S$ layers are of the same
thickness.
The vertical arrows mark the end, as  $H_ B$ decreases (top)
or $\Lambda$ increases (bottom, see inset)
of the region of stability of a certain state (see text).
}
\end{figure}

\begin{figure}[t]
\includegraphics[scale=.5,angle=0]{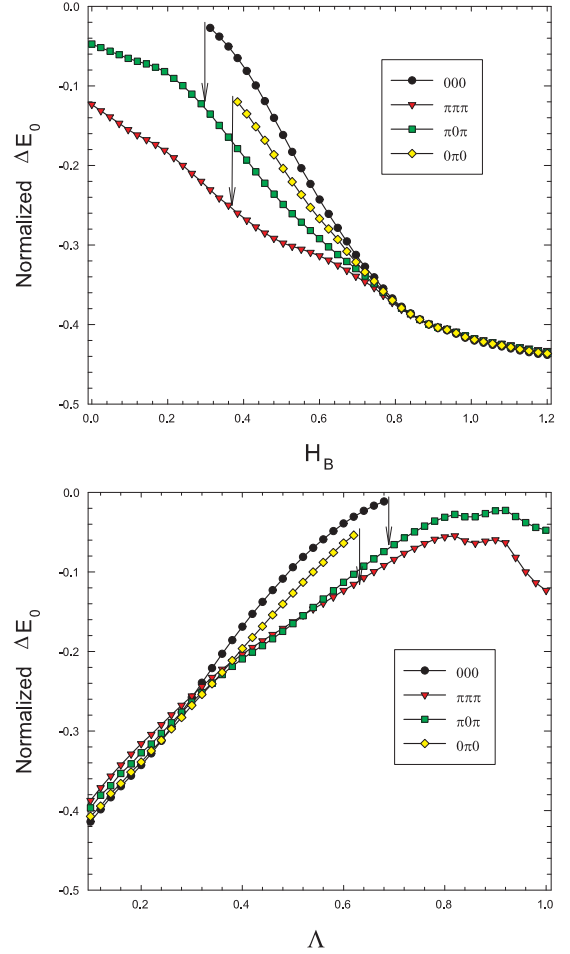}
\caption{\label{fig8} (Color online). Normalized $\Delta E_0$ for a seven layer
$SFSFSFS$ structure, as in Fig.~\ref{fig7}, but for the case
where the thickness of the two inner S layers is doubled. 
 }
\end{figure}

In Figure \ref{fig6} we plot $\Delta E_0$, defined and normalized
as explained, for a three layer
$SFS$ system. As in previous Figures, we have $D_S=100$, $D_F=10$
and $I=0.2$.  Results for self consistent states of both
the 0 and $\pi$ type are plotted as indicated. The top panel  shows
$\Delta E_0$ as a function
of the barrier thickness parameter $H_B$ at $\Lambda=1$. 
The bottom panel plots the same quantity as a  
function of mismatch $\Lambda$ at zero barrier and should be viewed 
in conjunction with
Fig.~\ref{fig2}. Looking first at the top panel, one sees that the zero
state is stable (has nonzero condensation energy) only for $H_B$ greater
than about $0.31$. 
An attempt to find a solution of the 0 type for $H_B$  just below
its ``critical'' value by using a 
solution of that type previously found for a slightly higher $H_B$
as the starting guess, 
and iterating the self consistent process, leads after many
iterations to a solution of the $\pi$
type. This is indicated
by the vertical arrow.  At larger
barrier heights, the two states become degenerate. This makes sense physically:
as the barriers become higher the proximity effect becomes
less important, and the S layers  behave more as independent
superconducting slabs. The relative phase is then immaterial. For even
larger $H_B$ we expect, from Eq.~\ref{bulke} and the geometry, the 
normalized quantity plotted to trend, from
above,  toward a limit $\approx -0.5(1-D_F/D_S) = -0.45$  and this is seen in the
top panel. One can also see that in the region of interest (barriers not too
high), the absolute value
of the condensation energy is substantially below that of the bulk.
In the bottom panel similar trends can be seen: in the absence of mismatch
($\Lambda=1$) only the $\pi$ state is found, and its condensation
energy exhibits a somewhat oscillatory
behavior as $\Lambda$ decreases from
unity. The 0 state does not appear
until $\Lambda$ is about $0.7$ and attempts (by the procedure just described)
to find it lead to a $\pi$ solution upon iteration (arrow).
This is in agreement with the results in 
Fig.~\ref{fig2}. For large mismatch the absolute value of $\Delta E_0$
increases, as the S slabs become more weakly coupled, with a trend  toward
the limiting value just discussed. A very important difference between
the top and bottom panels, however,
is the crossing of the curves near $\Lambda=0.33$ in Fig.~\ref{fig6}(b). This is
in effect a first order phase transition between the $\pi$ and 0 configurations
as the mismatch changes.     

The results of performing the same study for a seven layer system with
four $S$ layers can be seen
in the next two Figures. Figure \ref{fig7} corresponds to the case where all
the $S$ layers have the same thickness ($D_S=100$, and all other parameters also
as in the previous figures), while in Fig.~\ref{fig8} the thickness of the
two inner $S$ layers is doubled. 
Results for the four possible symmetric junction configurations 
mentioned in conjunction with Fig.~\ref{fig3} are 
given, as indicated in the legends of Figs.~\ref{fig7}
and \ref{fig8}. Three of those configurations, $000$, $\pi\pi\pi$ and
$\pi0\pi$ have appeared among the results
in Figs.~\ref{fig3} and \ref{fig4}.  The other
configuration corresponds to the $0 \pi 0$ sequence. 
\begin{figure}[t]
\includegraphics[scale=.4,angle=0]{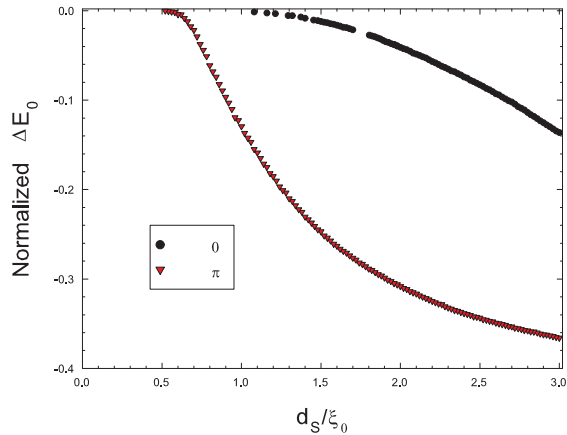}
\caption{\label{fig9} (Color online). The  normalized $\Delta E_0$ 
for a three
layer $SFS$ system, as a function of the thickness $d_S$ of the $S$ layers
(given in units of $\xi_0$) at fixed $D_F=10$, $I=0.2$ and without a barrier
or mismatch. Results  for 0 and $\pi$ self consistent
states are given as indicated. 
 }
\end{figure} 
We see that there are some striking differences between these examples
and the three layer system. While in the latter
case a configuration ceases to exist only
when its condensation energy tends  to zero, now configurations
can become unstable even when, for nearby values of 
the relevant parameter, they still have a negative condensation energy.
As this occurs, 
the
vertical arrows in each panel indicate (an inset is needed in one case
for clarity) how the states transform into each other as one varies the
parameters from the unstable to the stable region. Regardless of whether 
the inner layers are doubled or not, the tendency is for the innermost 
junction to remain 
of the same type, while the two outer junctions flip. 
Comparing  Fig.~\ref{fig7} and Fig.~\ref{fig8} we see that
the doubling of the inner layers has a clear quantitative effect without
having any strong qualitative influence. An important difference 
between the two cases is that in the first (all $S$ layer widths equal) the two
possible states ($\pi0\pi$ and $\pi\pi\pi$) at zero barrier and no mismatch
are nearly degenerate, while in the other case, the $\pi\pi\pi$ configuration
is favored. In the first case, the degeneracy
is removed as the barrier height begins to increase, but $\Lambda$ has a small
effect in relative stability. In Fig.~\ref{fig8}, bottom panel, the 
oscillatory effect of $\Lambda$ near the no-mismatch limit is seen, as
in the three layer case. For large mismatch or barrier, the results become
again degenerate and trend towards the appropriate limit. We expect these
seven layer results to be at least qualitatively 
representative of what occurs for larger values of $N_L$: thus, states
of the types
$000\cdots 000$, $\pi\pi\cdots \pi\pi$, and $\pi 00\cdots 00\pi$ (outer
junctions one way and inner ones the other) should predominate for large $N_L$.

\begin{figure}[t]
\includegraphics[scale=.7,angle=0]{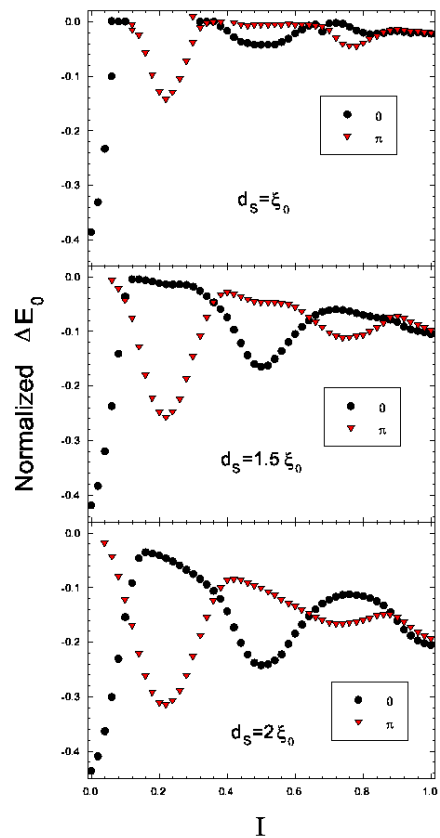}
\caption{\label{fig10} (Color online). The normalized $\Delta E_0$ 
for a three
layer $SFS$ junction, as a function of the parameter $I$,
for three different $S$ thicknesses (as labeled) and
fixed $D_F=10$.  Results  for 0 and $\pi$ self consistent 
states are given as indicated. 
 }
\end{figure} 

It is at least of equal interest to study how the stability depends on 
the geometry.
We discuss this question in the next four figures. First, in Fig.~\ref{fig9}
we present results for the condensation free energy of a three layer system as a 
function of $d_S/\xi_0$ at fixed $D_F=10$, $I=0.2$, $H_B=0$, $\Lambda=1$.
We see that $d_S$ must be at least half a correlation length for 
superconductivity to be possible at all in this system.
Convergence  near that value is rather slow, requiring approximately 200
iterations.
The superconducting state then begins
occurring,  for this value of $D_F$, in a $\pi$ configuration only.
When $d_S$ reaches $\xi_0$, the $\pi$ state condensation energy reaches
already an appreciable value that is consistent with that seen in the appropriate
limits of the panels in Fig.~\ref{fig6}. The $0$ state is still not
attainable (again, consistent with Fig.~\ref{fig6}) until $d_S$ reaches a
somewhat larger value. The condensation energies of the two states 
converge slowly toward each other 
upon increasing $d_S$,
but remain clearly non-degenerate
well beyond the range plotted.
The
small breaks in the $0$ state curve correspond to specific
$S$ widths that permit only the $\pi$   state.

\begin{figure}[t]
\includegraphics[scale=.55,angle=0]{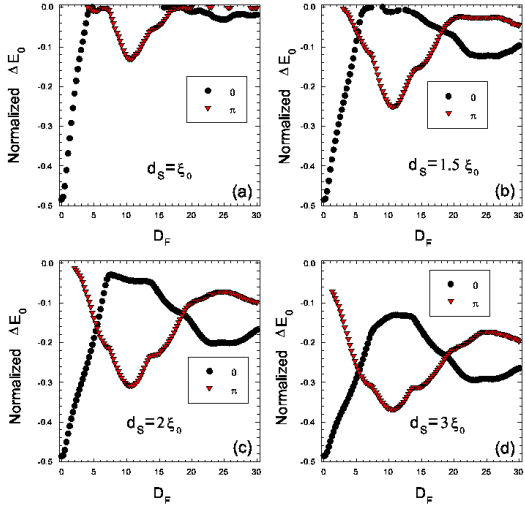}
\caption{\label{fig11} (Color online). The  normalized $\Delta E_0$ 
for a three
layer $SFS$ system, as a function of $d_F$, (rather than of $I$ as in
Fig.~\ref{fig10})
for four different $S$ thicknesses (as labeled) and
fixed $I=0.2$, $H_B=0$, $\Lambda=1$.  
 }
\end{figure} 

The behavior seen  in Fig.~\ref{fig9} depends strongly on $I$. This dependence
is displayed in Fig.~\ref{fig10} where we show the normalized
$\Delta E_0$ for
the same system, as a function of $I$, for three different values of $d_S$. 
For the value $I=0.2$, the results shown are consistent with Fig.~\ref{fig9},
including the nonexistence of the 0 state at $d_S=\xi_0$. We now see, however,
that it is not always the $\pi$ state which is favored, but that the difference
in condensation energies is an oscillatory function of $I$. This
of course reflects that whether the 0 or the $\pi$ state is preferred
depends, all other things being equal,  on the relation between $D_F$
and $(k_\uparrow-k_\downarrow)^{-1}$, and this
quantity (see Eq.~(\ref{xi2})) 
depends on $I$. At  intermediate values
of $I$ (centered around $I=0.5$) the zero state is favored, and when $I$ becomes
very small the $\pi$ state ceases to exist altogether (this can be seen also by examination
of Fig.~\ref{fig5}). As $I \rightarrow 0$ the condensation energy of the
0 state remains somewhat above the bulk value and, as one would expect, 
decreases slightly with increasing $d_S$. At larger values of $I$, the absolute
values of $\Delta E_0$ increase with $d_S$ and on the average decrease
slowly with $I$. 

\begin{figure}[t]
\includegraphics[scale=.60,angle=0]{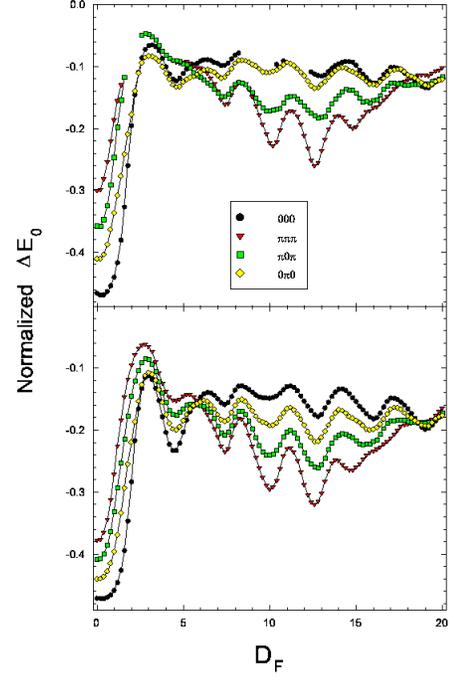}
\caption{\label{fig12} (Color online). The  normalized condensation 
free energy 
for a seven
layer $SFSFSFS$ system, as a function of $D_F$,
for 
fixed $I=0.2$, $d_S=\xi_0$, $\Lambda=1$
and $H_B=0.5$.  Results  for the four possible 
symmetric self consistent states are given, as indicated,
for both the cases where all $S$ layers are identical (top) and
where the thickness of the inner ones is doubled (bottom). 
Lines are guides to the eye. Breaks (top panel) indicate regions
where a certain configuration is not found. }
\end{figure} 

The oscillations in Fig.~\ref{fig10} as a function of $I$ at constant $d_F$
can also be displayed by considering results as a function of $d_F$ at
constant $I$. We do so in Fig.~\ref{fig11} where we plot, for
a three layer $SFS$ system,  $\Delta E_0$ as
a function of $D_F$ at constant $I=0.2$ for four values of $d_S/\xi_0$.
One sees again that for this value of $I$ the 0 state does not exist at 
$d_S=\xi_0$ and $D_F=10$ but that it appears at larger values of $d_S/\xi_0$.
The damped oscillatory
behavior is quite evident. At larger values of $d_F$ the condensation
energies of the two states trend towards a common value that increases in
absolute value with $d_S$. At a very small value of $d_F$, 
which depends on $d_S$, the $\pi$ state begins
to vanish, and the condensation free energy of the 0 state tends then towards 
the
bulk value. All of this is consistent with simple physical arguments.

In Fig.~\ref{fig12} we extend the results of Fig.~\ref{fig11} to the seven 
layer system. In this case we consider only one value of $d_S$  ($d_S=\xi_0$)
but include a finite barrier thickness, $H_B=0.5$.
The finite barrier allows for the possibility of more
distinct states coexisting (see Figs.~\ref{fig7}-\ref{fig8}).
We consider both the
cases where all $S$ layers thicknesses are equal (top panel) and the case
where the inner ones are doubled (bottom panel). All possible 
symmetric self consistent states 
were studied, as
indicated in the figure. In contrast with the three layer
example with no barrier, in the seven layer cases
with $H_B=0.5$ all of the four 
symmetric states ($000,\pi\pi\pi,\pi0\pi,0\pi0$)
are at least metastable over a range of  $d_F$, even
at $d_S=\xi_0$.
In the top panel we see however that only the $0\pi0$ state
is stable over the whole $d_F$ range. The $\pi\pi\pi$ 
state reverts to the $0\pi0$ state in the range $1.6\lesssim d_F\lesssim 4.2$,
while the 
$\pi0\pi$ state reverts to $000$ state for $1.8\lesssim d_F \lesssim 2.4$.
The $000$ state
is unstable for much of the range for $6\lesssim d_F \lesssim 12$. 
It appears that
in this range the $000$ state is sufficiently close
to a crossover (see e.g. Fig.~\ref{fig7}) that attempts to find it sometimes
converge to an asymmetric $\pi 0 0$ state, 
rather than the expected $\pi0\pi$. In these cases
the number of iterations to convergence is substantially increased,
as the order parameter attempts to readjust its profile.
For the situation where the inner $S$ layers are twice the width
of the outer ones, we see (bottom panel) that all
four symmetric configurations are either stable or
metastable for the whole $d_F$ range. This is consistent
with Fig.~\ref{fig7}, where at $H_B=0.5$ all four states
are present simultaneously. The condensation energy
is of course lower than in the previous case, due to the
increased pairing correlations associated with the thicker $S$
slabs. For both geometries oscillations 
arising from the scattering potential lead to deviations from
the estimated periodicity determined by $(k_\uparrow-k_\downarrow)^{-1}$.
For sufficiently large $d_F$ the difference in energies becomes small. 
One can infer from these results than in superlattices
with realistic oxide barriers,
where as the number of layers increases a larger number
of nontrivial possible states arise, the 
number of local free energy
minima that can coexist will increase.

\subsection{DOS}

\begin{figure}[t]
\includegraphics[scale=.65,angle=0]{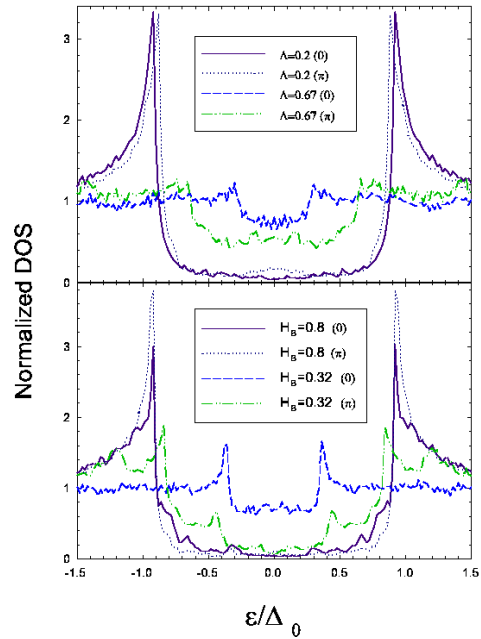}
\caption{\label{fig13} (Color online). The normalized DOS (see
text) for a  $SFS$ trilayer, plotted as
a function of the energy (in units of the bulk zero temperature gap).
Results for both 0 and $\pi$ self consistent
states are given, as indicated.
In the top panel, the DOS is shown at $H_B=0$ 
for two different values of the mismatch
parameter, $\Lambda=0.2$, and 0.67,
the latter being a case for which the 0 state is nearly unstable
(see  Fig.~\ref{fig6}).
The bottom panel shows the DOS profile in the absence of mismatch 
($\Lambda=1$), but with
the interface scattering parameter $H_B$ taking on the two values shown,
chosen on similar criteria as the $\Lambda$ values (see text).
 }
\end{figure} 

\begin{figure}[t]
\includegraphics[scale=.8,angle=0]{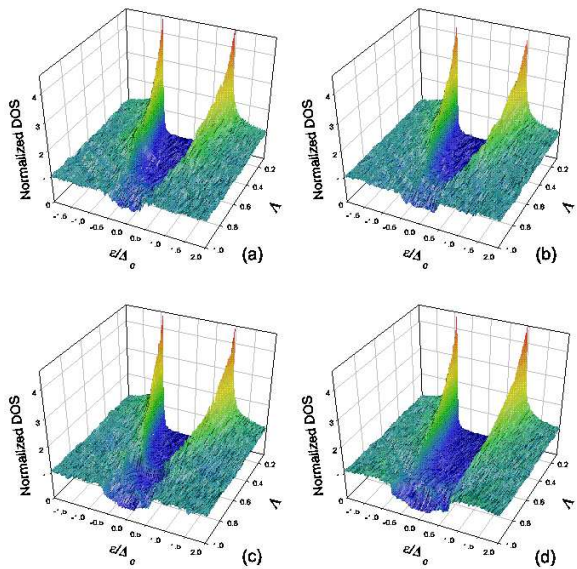}
\caption{\label{fig14} (Color online). The normalized local DOS  for a seven
layer system versus the dimensionless energy and mismatch parameter 
$\Lambda$. The panels are arranged
as those in Fig.~\ref{fig3}: (a) and (b) show results obtained
when all $S$ layers have the same thickness, while in (c) and (d)
that thickness is doubled. Panels (a) and (c) correspond to  
starting the iteration procedure with a $000$ order parameter, 
while panels (b) and (d) are
obtained by starting with  a $\pi\pi\pi$ junction, in which case
the self-consistent solution is also always of this type. 
}
\end{figure} 

We conclude this section by presenting some results for the DOS, as it can be
experimentally measured. 
The results given are in all cases for the quantity ${N}_\sigma(z,\epsilon)$
defined in Eq.~(\ref{dos}), summed over $\sigma$, and 
integrated over 
a distance of one coherence length from the edge of the sample.
We  normalize our results to the
corresponding value for the normal state of the superconducting material,
and the energies to the bulk value of the gap, $\Delta_0$.
All  parameters not otherwise mentioned
are set to their ``default'' values outlined
at the beginning of this section.

In the top panel of Fig.~\ref{fig13} we present results for an $SFS$ 
trilayer, for
two contrasting values of the mismatch parameter $\Lambda$. 
As usual, results labeled as ``0"  and  ``$\pi$" are for the
case where the self consistent states plotted are of these types.
The $0$ and $\pi$ state curves corresponding to $\Lambda=0.67$, where
(see Fig.~\ref{fig6}) the 0 state is barely metastable,
have clearly distinct signatures, with a smaller  gap opening for the $0$ 
state, and consequently
more subgap quasiparticle states.
Therefore, as can be seen in  Fig.~\ref{fig2}, 
when there is little
mismatch the pair amplitude is relatively large
in the $F$ layer. 
When there is more mismatch the
proximity effect weakens, the gap opens and large peaks form which 
progressively become more BCS-like as the mismatch increases ($\Lambda$
decreases) to $\Lambda=0.2$. This progression takes a different form for the 0
and $\pi$ states, as one can see by careful comparison of the curves. 
In the bottom panel we demonstrate the effect
of the barrier height. 
The results are displayed as in the top panel, but with the dimensionless
height $H_B$ taking the place of the mismatch parameter. This Figure
should be viewed in conjunction with the top panel
of Fig.~\ref{fig6}. One of the values
of $H_B$  chosen ($H_B=0.32$) is again such that the
0 state is barely metastable, while for the other value
the 0 and $\pi$ states  have similar condensation energies.
For
$H_B$ close to 0.31 there is a marked contrast between the two
plots, with the gap clearly opening  wider, and containing more
structure for the $\pi$ state. 
At larger $H_B$ the gap becomes larger in both cases, with
the BCS-like peaks increasing in  height.
Thus, as the barriers becomes larger, 
one is dealing with nearly independent superconducting slabs
and the plots become more similar. The largest difference therefore, occurs,
as for the mismatch, at the intermediate values more likely to be found
experimentally.

It is also illustrative to display the DOS
in a three dimensional format. 
Thus, in Fig.~\ref{fig14}, the normalized DOS
is presented as a function of  $\epsilon/\Delta_0$ and $\Lambda$ 
for the seven layer
structures previously discussed. 
This permits a much more extensive range of $\Lambda$
values to be examined.
The panel arrangement is the same as in
Fig.~\ref{fig3}: in the top two panels, all $S$ layers have the same
thickness ($d_S=\xi_0$) while in the bottom ones the thickness of the inner
ones is doubled. On the right column (panels (b) and  (d)) the 
self-consistent results shown were obtained by starting the iteration
process with an initial guess of the $\pi\pi\pi$ type,
in which case, as one can see from Fig.~\ref{fig3} and the bottom
panels of Figs.~\ref{fig7} and \ref{fig8}, the solution is also of the
$\pi\pi\pi$ type. The left column panels were obtained from an initial
guess of the $000$ type  and therefore correspond, as
one can see from Figs.~\ref{fig3}, \ref{fig7} and \ref{fig8}, to solutions
of the $000$ type for $\Lambda \lesssim 0.5$ (panel (a)) and $\Lambda \lesssim 0.7$
(panel (c)), and to solutions of the $0\pi0$ type otherwise. 
There is no overlap among
the results in the four panels. One sees again that as mismatch increases
the BCS like peaks become more prominent and move out, trending towards
their bulk positions, while the gap opens.
Furthermore, the $\pi\pi\pi$ results for the doubled $d_S$ inner layers
are smoother and show clearly a broader gap throughout, due to the
extended $S$ geometry.
The DOS behavior observed here is entirely consistent with the
condensation energy results found.

\section{conclusions \label{conclusions}}

In summary, we have found self consistent solutions to the microscopic
BdG equations for $SFS$ and $SFSFSFS$ structures, for a wide range
of parameter values. We have shown that, in most cases, several such
self consistent solutions 
coexist, with differing spatial
dependence of the pair potential $\Delta({\bf r})$ and the pair amplitude $F({\bf r})$.
Thus, there can be in general competing local minima of the free energy.
Determining their relative stability requires the 
computation of their respective condensation free energies,
which we have done by using an efficient, accurate approach
that does not involve the quasiparticle amplitudes directly, and requires only
the eigenenergies and the pair potential. 

For $SFS$ trilayers (single junctions), we found that
both $\pi$ and $0$
junction states exist for a range of values of the relevant
parameters.
We have displayed  results for the pair amplitude, which give insight into the
superconducting correlations, and for the 
condensation free energies 
of each configuration, to determine the true equilibrium state.
We have shown that a transition (which is in effect
of first order in parameter space) can occur between the $0$ and $\pi$ 
states
for a critical value of $\Lambda$. 
The difference in condensation energies between the two possible states
exhibits oscillations as a function of $I$. This
behavior is strongly dependent on
the width of the $S$ layers. For $d_S$ equal to one coherence length $\xi_0$, 
there exists a range of $I$ in which either a $0$ or $\pi$ state survives,
but not both. Increasing  the $S$ width by about $\xi_0/2$ restores
the coexistence of both states.

Several interesting phenomena
arise when one explores the geometrical parameters of trilayer structures.
For a fixed ferromagnet width $d_F$, and parameters values that lead to
a $\pi$ state, we found (see Fig.~\ref{fig9}) that the $\pi$ configuration 
remains the ground state
of the system as $d_S$ varies.
The $\pi$ state first appears
at small $d_S$ ($d_S \approx \xi_0/2$), and then its condensation energy
declines monotonically towards the bulk limit. The 
metastable $0$ state begins 
at larger $d_S\approx \xi_0$, and its condensation
energy  declines also slowly over the range of $d_S$ studied.
The other relevant length that was considered is $d_F$.
The condensation energy is, for both states, an oscillatory function
of $d_F$. The oscillations become better defined, and
the possibility of both $0$ and $\pi$ states coexisting increases,
at larger $d_S$. 
As expected, we find 
that the condensation energy has very similar properties when either
$d_F$ or 
$I$ varies. The period approximately agrees  with the estimate 
given by $(k_\uparrow-k_\downarrow)^{-1}$, which
governs the oscillations of the pair amplitude and in general, of other
single particle quantities.
The results for the DOS reflect the crossover from a
state with populated subgap peaks to
a nearly gapped BCS-like behavior
as $\Lambda$  is decreased or the barrier height $H_B$ increased. 
Signatures that may help to experimentally identify the $0$ and $\pi$
configurations
were seen.

As the number of layers increases, so does the number of competing
stable and metastable junction configurations.
We considered two types of seven layers structures, and found
that doubling the width of the  inner $S$ layers (which
are bounded on each side by ferromagnets), resulted typically in
different quasiparticle spectra and
pair amplitudes, compared to the situation when all $S$ layers have
the same $d_S$. 
For large mismatch or barrier strength, the phase of the
pair amplitude in each layer
is independent, and configurations are nearly degenerate,
but as each of these parameters diminishes
there is a crossing
over to a situation where the free energies
of each configuration are well-separated. At 
certain values of $\Lambda$ and $H_B$, some
configurations become unstable.
These values are different depending on
the type of system (single or double inner layers).
For fixed $I$ and $d_F$ we found
that if a
state is stable at no mismatch and zero barrier,
then it remains at least metastable over a very wide range of $\Lambda$
and $H_B$ values. 
Our results showed that
self-consistency  cannot be neglected as the number of layers increases, 
due to the nontrivial  and intricate spatial variations in
$F({\bf r})$ that become possible.

For seven layers, we studied in detail the 
condensation free energies of the four symmetric junction states,
$000,\pi\pi\pi,\pi0\pi$ and $0\pi0$, in the previously
introduced notation. 
We first investigated the stable states as a function of $\Lambda$ and $H_B$.
In contrast to the three layer system,
we found that  states could become unstable even 
when the condensation energy did not tend to zero  for nearby
values of the relevant parameters.
For double width inner layered structures, we found a greater spread in the 
free energies
between the four states, and the instability found in
certain cases for the $000$ and $0\pi0$ states
was shifted in $\Lambda$ and $H_B$, in agreement with the pair amplitude
results. It is reasonable to assume that these results
are representative of what occurs for
superlattices. We again found  transitions upon varying $\Lambda$ and the
number and sequence of the transitions is now more intricate (see
Figs.~\ref{fig7} and \ref{fig8}).
The analysis of the geometrical properties 
revealed that scattering
at the interfaces  modifies the expected damped oscillatory
behavior of the condensation energy as a function of $d_F$. In effect, the 
barriers introduce significant atomic scale oscillations that smear 
the periodicity. This underlines the importance of a microscopic approach for
the investigation of nanostrucutres.
As with the $\Lambda$ dependence, we also showed that
the global minima in the free energy is different for the two
structures as $d_F$ changes. 
The configuration of the ground state of the system 
with $S$ layers of uniform width was more variable in parameter space,
compared to when the inner layers are doubled, (compare Figs.~\ref{fig7}
and \ref{fig8}).
Finally, we calculated the  DOS, to illustrate and compare the 
differences in the spectra for the two different seven layer geometries.
Of the two, the energy gap for the single inner layer case,
and $\pi\pi\pi$ stable state,
contains more subgap states, for the specific examples
plotted (Figs.~\ref{fig13} and \ref{fig14}) due to stronger pair 
breaking effects
of the exchange field.
These states fill in increasingly with decreased mismatch.

Our results were obtained in the clean limit, which is appropriate for
the relatively thin structures envisioned here. Furthermore, as shown in
Ref.~\onlinecite{hv2} in conjunction
with realistic comparison with experiments,\cite{mcp} the influence of impurities can be taken
into account by replacing the clean value of $\xi_0$ with an effective
one. A separate important issue is that of the free energy barriers
separating the different free energy minima we have found, and hence
to which degree are metastable states long lived. Our method 
cannot  directly  answer this question, but from the macroscopic
symmetry  differences  in the pair amplitude structure
of the different states one would have to conclude that the barriers are
high and the metastable states could be very long lived. We expect
that the transitions found here in parameter space at constant
temperature will be reflected
in actual first order phase transitions as a function of $T$.  Such transitions would presumably be very
hysteretic. We hope
to examine this question in the future.






\begin{acknowledgments}
The work of K.~H. was supported in part by a 
grant of HPC time from the Arctic Region Supercomputing Center
and by ONR Independent Laboratory In-house Research funds.
\end{acknowledgments}


\end{document}